\documentclass[11pt]{article}
\usepackage{geometry}                
\usepackage{graphicx}
\usepackage{amssymb}
\usepackage{epstopdf}
\usepackage{epsfig}
\usepackage{hyperref}

\title{Neutrino Astrophysics and Galactic Cosmic Ray Anisotropy in IceCube}

\author{P. DESIATI$^*$ for the IceCube Collaboration$^\dag$\\
IceCube Research Center, University of Wisconsin,\\
Madison, WI 53703, U.S.A.\\
$^*$E-mail: {\tt desiati@icecube.wisc.edu}\\
$^\dag$ {\tt http://icecube.wisc.edu}}

\date{}

\begin{document}
\maketitle

\begin{abstract}
The IceCube Observatory is a kilometer-cube neutrino telescope under construction at the South Pole and planned to be completed in early 2011. When completed it will consist of 5,160 Digital Optical Modules (DOMs) which detect Cherenkov radiation from the charged particles produced in neutrino interactions and by cosmic ray initiated atmospheric showers. IceCube construction is currently 90\% complete.  A selection of the most recent scientific results are shown here. The measurement of the anisotropy in arrival direction of galactic cosmic rays will also be presented and discussed.
\end{abstract}


\section{Introduction}
\label{sec:intro}

The IceCube Observatory is a km$^3$ neutrino telescope designed to detect astrophysical neutrinos with energy above 100 GeV. IceCube observes the Cherenkov radiation from charged particles produced in neutrino interactions.

The quest for understanding the mechanisms that shape the high energy Universe is taking many paths.  Gamma ray astronomy is providing a series of prolific experimental observations, such the detection of TeV $\gamma$ rays from point-like and extended sources, along with their correlation to observations at other wavelengths. These observations hold the clues about the origin of cosmic rays and the possible connection to shock acceleration in Supernova Remnants (SNR), Active Galactic Nuclei (AGN) or Gamma Ray Bursts (GRB). Supernov{\ae} are believed to be the sources of galactic cosmic rays, nevertheless the $\gamma$ ray observations from SNR still do not provide us with a definite and direct evidence of proton acceleration. The competing inverse Compton scattering of directly accelerated electrons may significantly contribute to the observed $\gamma$ ray fluxes, provided that the magnetic field in the acceleration region does not exceed 10 $\mu$G \cite{abdo}.

Ultra High Energy Cosmic Rays (UHECR) astronomy, has the potential to hold the key to a breakthrough in astroparticle physics. The identification of sources of cosmic rays could provide a unique opportunity to probe the hadronic acceleration models currently hypothesized. On the other hand, cosmic ray astronomy is only possible at energies in excess of 10$^{19}$ eV, where the cosmic rays are believed to be extragalactic and point back to their sources. TeV $\gamma$ rays from those sources are likely absorbed during their propagation between the source and the observer : at $\sim$10 TeV $\gamma$ rays have a propagation length of about 100 Mpc, while at $\sim$100 GeV $\gamma$ rays can propagate much deeper through the Universe.

If the extra-galactic sources of UHECR are the same as the sources of $\gamma$ rays, then hadronic acceleration is the underlying mechanism and high energy neutrinos are produced by charged pion decays as well. Neutrinos would provide an unambiguous evidence for hadronic acceleration in both galactic and extragalactic sources, and are the ideal cosmic messengers, since they can propagate through the Universe undeflected and with practically no absorption. But the same reason that makes neutrinos ideal messengers makes them also difficult to detect.

The discovery of the anisotropy in arrival direction of the galactic cosmic rays has also triggered particular attention recently.
The origin of galactic cosmic ray anisotropy is still unknown. The structure of the local interstellar magnetic field within 1 pc is likely to have an important role in shaping the large angular scale features of the observed anisotropy. Nevertheless it is possible to argue that the anisotropy might be originated by a combination of astrophysical phenomena, such as the distribution of nearby recent supernova explosions \cite{erlykin}. The observation of galactic cosmic ray anisotropy at different energy and angular scales has, therefore, the potential to reveal the connection between cosmic rays and shock acceleration in supernovae.

At the same time, there seems to be clear observational evidence for the existence of dark matter in the Universe, even if its nature remains unknown. A variety of models predict the existence of a class of non-relativistic particles called Weakly Interacting Massive Particles (WIMP). These particles could be gravitationally condensed within dense regions of matter (such as the Sun or the galactic halo) and could provide a visible source for indirect detection via neutrino generation through annihilation. Neutrino telescopes are powerful tools to indirectly test spin-dependent WIMP-nucleon scattering cross section, provided the models for matter distribution and WIMP annihilation rate are taken into account.

In section \S \ref{sec:ic} the IceCube Observatory apparatus functionality and calibration are described. Selected physics analyses results are summarized in \S \ref{sec:phys} : the determination of the atmospheric muon neutrino energy spectrum (\S \ref{ssec:atm}), the search for astrophysical neutrinos from diffused and point sources, and from Gamma Ray Bursts (\S \ref{ssec:astro}), the indirect search for dark matter (\S \ref{ssec:dm}), and the anisotropy in cosmic rays arrival direction (\S \ref{ssec:anyso}). 

\section{The IceCube Observatory}
\label{sec:ic}

\begin{figure}[h]
\begin{center}
\psfig{file=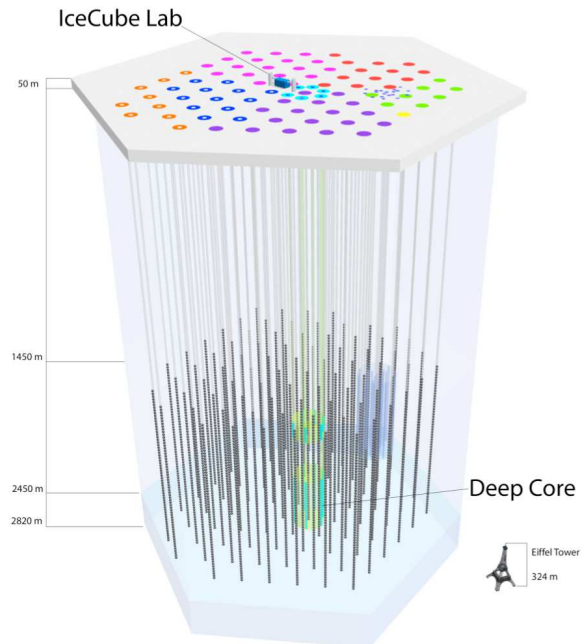,width=3in}
\caption{The IceCube Observatory. Currently it consists of 79 strings and 4,740 DOMs. In 2011 it will be completed with 86 strings and 5,160 DOMs. The shaded region near the center of the array is the Deep Core dense sub-array, and the one on the right is the AMANDA neutrino telescope, de-comissioned in May 2009.}
\label{fig:icecube}
\end{center}
\end{figure}

The IceCube Observatory (see Fig. \ref{fig:icecube}) currently consists of 4,740 DOMs deployed in 79 vertical strings (60 DOMs per string) between 1,450 m and 2,450 m depth below the Geographic South Pole. At the beginning of 2011 IceCube will be completed with 86 strings and 5,160 DOMs. The surface array IceTop with 81 stations, each consisting of two tanks with frozen clean water with two DOMs each, will provide the measurement of the spectrum and mass composition of cosmic rays at the knee and up to about 10$^{18}$ eV. The Deep Core sub-array, consisting of 6 densely instrumented strings and located at the bottom-center of IceCube, is capable of pushing the neutrino energy threshold to about 10 GeV. The surrounding IceCube instrumented volume can be used to veto the background of cosmic ray induced through-going muon bundles to enhance the detection of down-going neutrinos within the Deep Core volume. The veto rejection power can reach 10$^5$.

\vskip 0.2cm

AMANDA was the first and largest neutrino telescope before the construction of IceCube. Using analog technology it significantly contributed to the advance of neutrino astrophysics searches and it was de-commissioned in May 2009.

The basic detection component of IceCube is the DOM : it hosts a 10-inch Hamamatsu photomultiplier tube (PMT) and its own data acquisition circuitry enclosed in a pressure-resistant glass sphere, making it an autonomous data collection unit. The DOMs detect, digitize and timestamp the signals from optical Cherenkov radiation photons. Their main board data acquisition (DAQ) is connected to the central DAQ in the IceCube Laboratory at the surface, where the global trigger is determined \cite{dom}. 

The detector calibration is one of the major efforts aimed at characterizing its response and to reduce systematic uncertainties at the physics analysis level. Each PMT is tested in order to characterize its response and to measure the voltage yielding a specific gain \cite{pmt}. In the operating neutrino telescope the current gain is about 10$^7$ and the corresponding dark noise rate is about 500 Hz. Time calibration is maintained throughout the array by regular transmission to the DOMs of precisely timed analog signals, synchronized to a central GPS-disciplined clock. This procedure has a resolution of less than 2 nsec. The LEDs on the flasher boards instrumented in the DOMs, are used to measure the photo-electron (p.e.) transit time in the PMT for the reception of large light pulses between neighboring DOMs. This delay time is given by light travel time from the emitter to the receiver, by light scattering in ice and by the electronics signal processing. The RMS of this delay is also less than 2 nsec. Waveform sampling amplitude and time binning calibration is periodically performed in each DOM and used to extract the number of detected p.e. with an uncertainty of less than 10\%. Higher level calibrations are meant to correlate the number of detected p.e. to the energy of physics events that trigger the detector. Instrumented devices, such as the flasher boards, are used to illuminate the detector with 400 nm wavelength photons (corresponding to the wavelength yielding the highest detection sensitivity), simulating a real electron-neutrino interaction, or cascade, inside the detector. A complete Monte Carlo simulation chain is used to relate the known number of injected photons with the energy scale of the artificial cascade. The energy resolution depends on the event topology (track-like versus cascade-like) and on its containment inside the instrumented volume. Monte Carlo simulations provide the necessary fluctuations implicit by the topology and containment of the physics events. The ice optical properties are the most fundamental calibration determination that allows us to know how photons propagate through the ice and, therefore, how to relate the number of detected p.e. to the energy of the physics events. Due to the antarctic glaciological history, the optical properties depend on depth and they have been measured in the past using AMANDA in-situ calibration lasers \cite{amanda}, in the depth range between 1,400 m and 2,000 m. The optical properties down to 2,450 m, the depth of IceCube instrumentation, are extrapolated from ice core observations in other location of the antarctic continent, and a new campaign of extended in-situ measurements is currently being carried out.

\section{Physics Results}
\label{sec:phys}

If the DOMs that detect Cherenkov photons satisfy specific trigger conditions, an event is formed and recorded by the surface DAQ. An on-line data filtering at the South Pole reduces the event volume to about 10\% of the trigger rate, based on a series of filter algorithms aimed to select events based on directionality, topology and energy. The filter allows us to transfer data via satellite to the northern hemisphere for prompt physics analyses.

\subsection{Atmospheric neutrinos}
\label{ssec:atm}

99.999\% of the events that trigger IceCube are muons produced by the impact of primary cosmic rays in the atmosphere. Only a small fraction of the detected events ($\sim$10$^{-5}$) are muon events produced by atmospheric neutrinos. In order to reject all the down-ward muon bundle background only up-ward events are generally selected, assuming specific event selections provide well reconstructed events. In the 40 string instrumented IceCube (IceCube-40) about 30-40\% of the up-ward events survive the selection, with a background contamination of less than 1\% \cite{warren}.

\vskip 0.2cm
\begin{figure}[h]
\begin{center}
\psfig{file=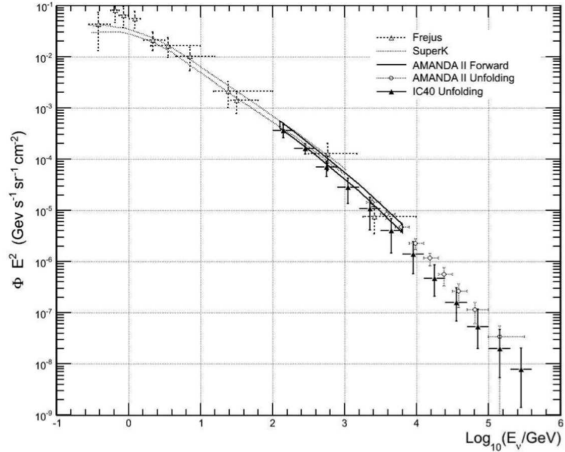,width=3in}
\caption{The preliminary unfolded spectrum of atmospheric muon neutrinos ($\nu_{\mu}+\bar{\nu}_{\mu}$) from the IceCube-40 detector \cite{warren}, compared with previously measurements : the Fr\'ejus result \cite{frejus}, upper and lower bands from Super-Kamiokande \cite{sk}, the forward-folding result from AMANDA \cite{john} and the unfolded spectrum from AMANDA \cite{kirsten}. The AMANDA unfolding analysis was a measurement of the zenith-averaged flux from 100$^{\circ}$ to 180$^{\circ}$ and the present analysis (IC40 unfolding) is a measurement of the zenith-averaged flux 
from 97$^{\circ}$ to 180$^{\circ}$.}
\label{fig:atmo}
\end{center}
\end{figure}

The energy estimation resolution of these atmospheric $\nu_{\mu}$ induced events is of the order of 0.3 in Log of neutrino energy, and a regularized unfolding technique is used to determine the energy spectrum. Fig. \ref{fig:atmo} shows the preliminary unfolded energy spectrum of the 17,682 atmospheric neutrinos detected by IceCube-40 between zenith angle of  97$^{\circ}$ and 180$^{\circ}$. The figure also shows other measurements performed by other experiments, including AMANDA \cite{john, kirsten}. IceCube has detected the highest energy atmospheric neutrinos (about 250 TeV where a significant fraction of neutrinos are expected to arise from the decay of heavy mesons with charm quarks). IceCube allowed us to extend the global measured spectrum up to 6 orders of magnitudes in energy. For the first time the precision of this measurement is providing a powerful tool to test the high energy hadronic interaction models that govern our present knowledge of the cosmic ray induced extensive air showers.

\subsection{Search for astrophysical neutrinos}
\label{ssec:astro}

Atmospheric neutrinos represent an irreducible background for the search of high energy astrophysical neutrinos. If hadronic acceleration is the underlying source of high energy cosmic rays and $\gamma$ rays observations, we expect that unresolved sources of cosmic rays over cosmological times have also produced enough neutrinos to be detected as a diffuse flux. Since shock acceleration is expected to provide an $\sim$E$^{-2}$ energy spectrum, harder than the $\sim$E$^{-3.7}$ of the atmospheric neutrinos, the diffuse flux is expected to dominate at high energy.

\vskip 0.2cm
\begin{figure}[h]
\begin{center}
\psfig{file=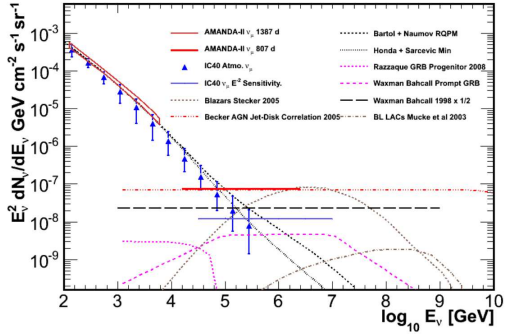,width=3in}
\caption{Current status on the search of diffuse astrophysical $\nu_{\mu}+\bar{\nu}_{\mu}$. The atmospheric neutrino background as measured by IceCube-40 and AMANDA are shown, along with the predictions to the highest possible energies \cite{bartol,naumov,honda,sarcevic}. The limit on an E$^{-2}$ flux by AMANDA \cite{diffama} and the preliminary sensitivity by IceCube-40 are shown. We predict that IceCube-40 sensitivity is below the Waxman \& Bahcall diffuse neutrino bound \cite{wb}. The graph shows other models of astrophysical muon neutrinos as well \cite{stecker, becker, becker2, razzaque, mucke}  (An E$^{-2}$ spectrum is a horizontal line in this graph).}
\label{fig:diff}
\end{center}
\end{figure}

Fig. \ref{fig:diff} shows the AMANDA experimental upper limit and the preliminary IceCube-40 sensitivity for an E$^{-2}$ diffuse muon neutrino spectrum ($\nu_{\mu}+\bar{\nu}_{\mu}$). One year of IceCube-40 is about 5 times more sensitive than 3 years of AMANDA and its sensitivity is lower than the Waxman-Bahcall neutrino bound. This means that IceCube is potentially approaching the discovery of the origin of cosmic rays.

In the Ultra High Energy range (i.e. above $\sim$10$^6$ GeV or UHE) IceCube is placing upper limits that are still more than about one order of magnitude above the predicted flux of neutrinos from UHECR interactions on the microwave photons (or GZK neutrinos \cite{gzk}). The complete IceCube Observatory might be able to reach the discovery level within the next 5-8 years.

If the observed $\gamma$ rays from galactic and extra-galactic point and extended sources are from neutral pion decays in hadronic acceleration sites or from cosmic ray interaction with molecular clouds, the charged pions could produce enough neutrinos that can be detected by a km$^3$ neutrino telescope.

\vskip 0.2cm
\begin{figure}[h]
\begin{center}
\psfig{file=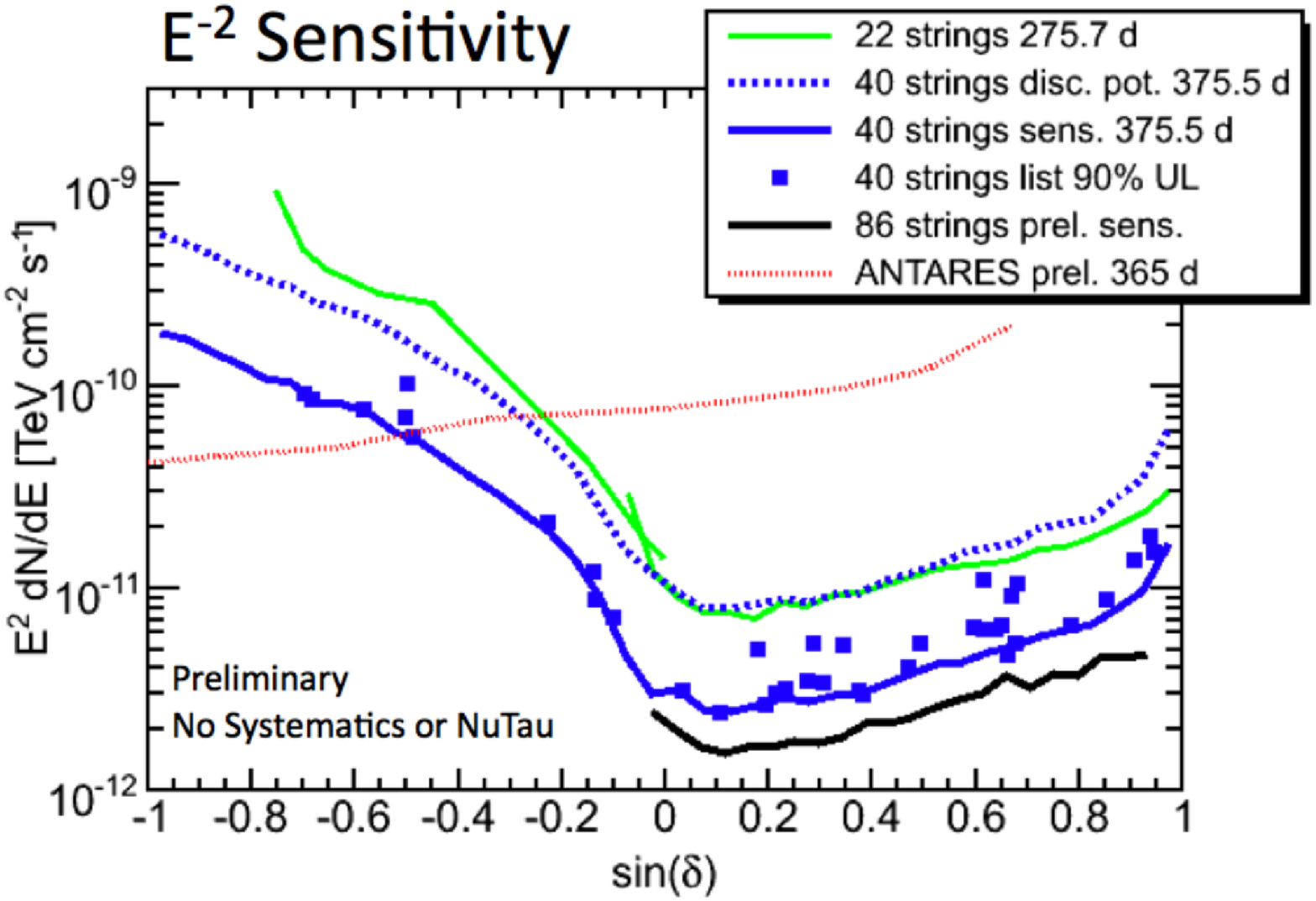,width=2.3in}
\psfig{file=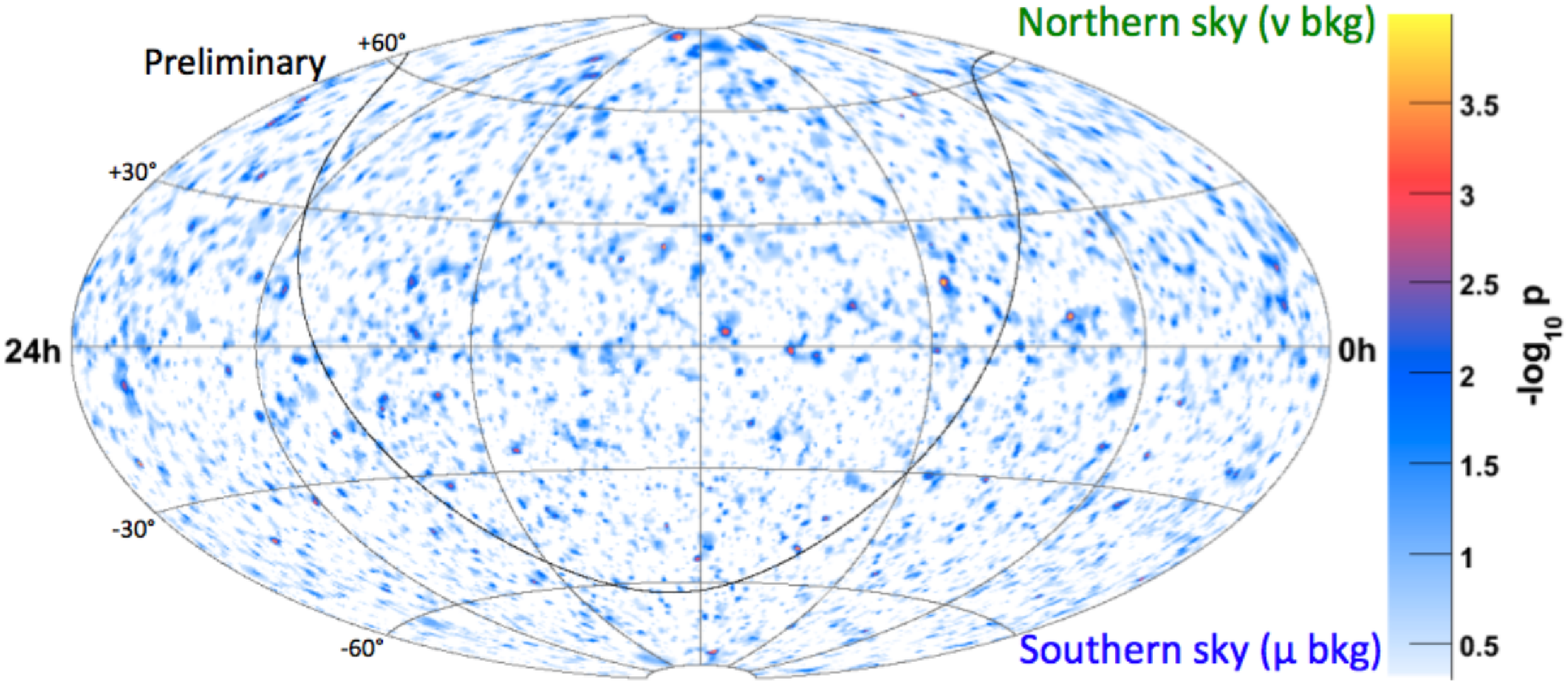,width=2.8in}
\caption{{\it On the left}: sensitivity (90\% CL) for a full-sky search of steady point sources of muon neutrinos with an E$^{-2}$ energy spectrum as a function of declination angle ($\delta<$ 0$^{\circ}$ for down-ward events, or the southern hemisphere, and $\delta>$0$^{\circ}$ for up-ward events, or the northern hemisphere). The IceCube-22 sensitivity for the full-sky search \cite{ic22south, ic22north} is about the same as the IceCube-40 discovery potential (5$\sigma$ discovery for 50\% of the trials), and about a factor of three larger than the IceCube-40 sensitivity (90\% CL). The squares show the IceCube-40 90\% upper limits for selected sources, and the black line represents the projected IceCube-86 sensitivity.
{\it On the right}: preliminary sky map in equatorial coordinates of statistical significance (p-value) from the search for steady point sources of high energy muon neutrinos in IceCube-40. The observation is extended to the southern hemisphere by reducing muon background five orders of magnitudes with a zenith-dependent energy selection ($>$100's TeV).}
\label{fig:point}
\end{center}
\end{figure}

\begin{figure}[h]
\begin{center}
\psfig{file=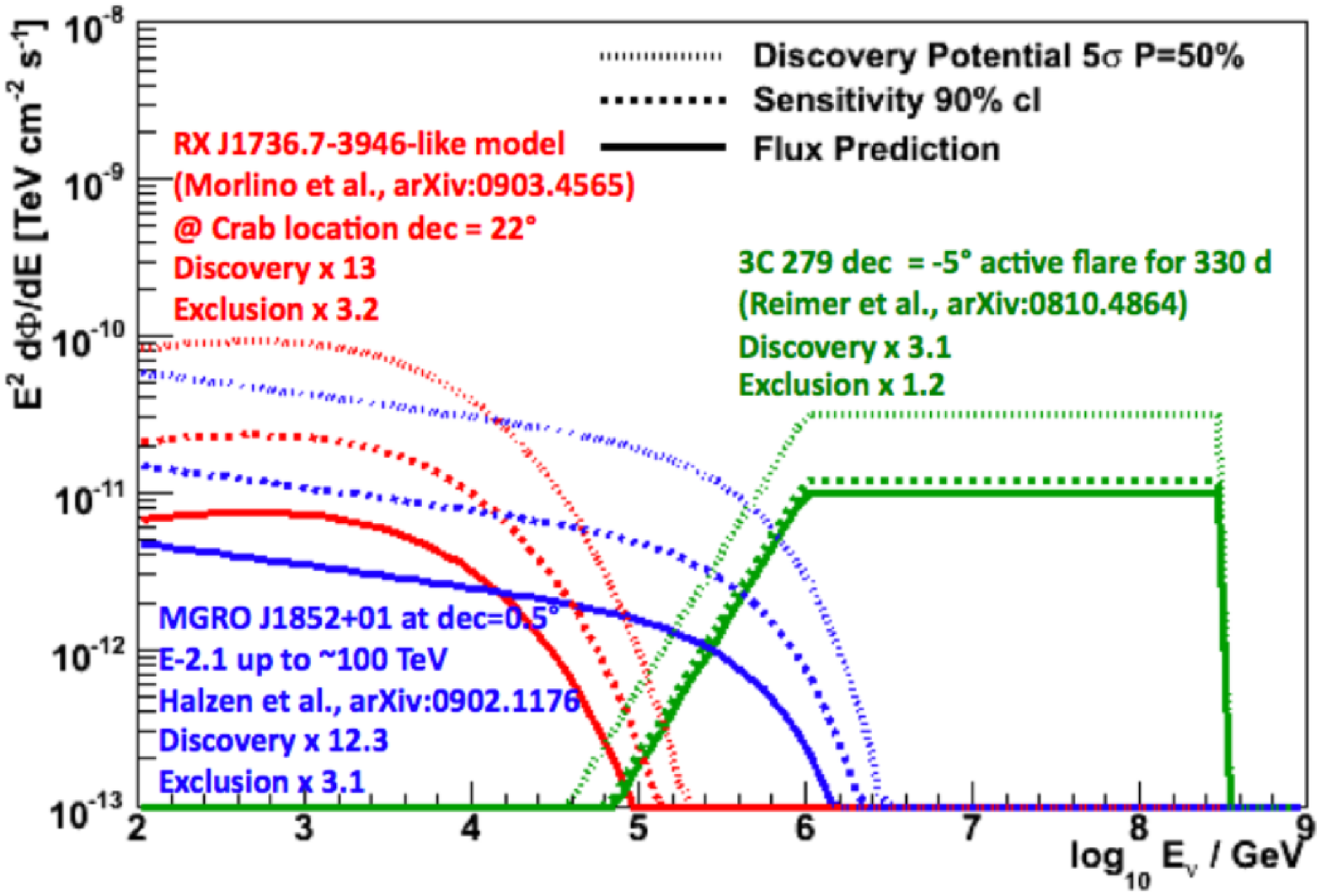,width=2.5in}
\psfig{file=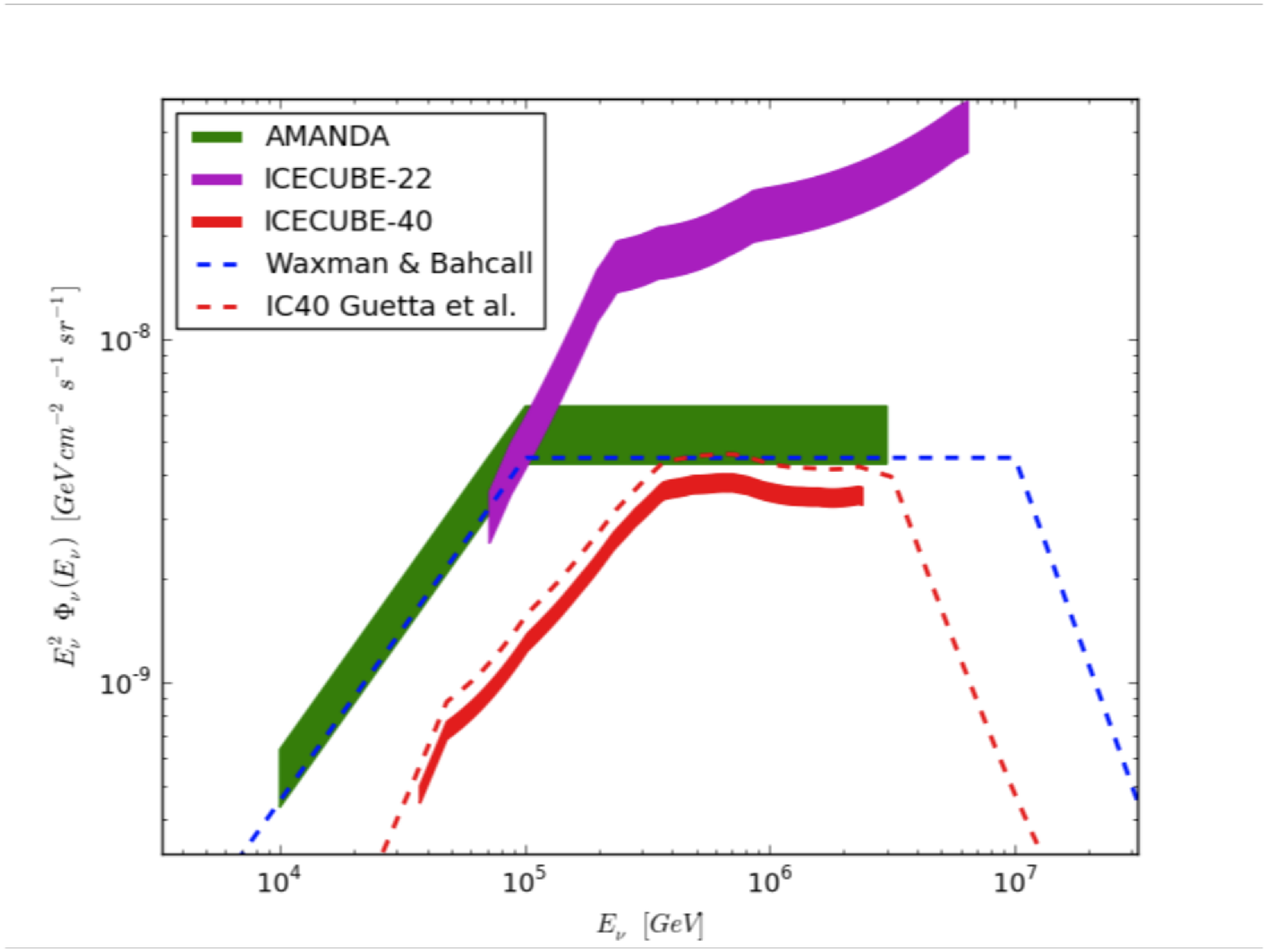,width=2.5in}
\caption{{\it On the left}: sensitivity (90\% CL) and discovery potential (5$\sigma$ discovery for 50\% of the trials) in IceCube-40 for the Crab \cite{crab}, MGROÊJ1852+01 \cite{milagro} and 3CÊ279 \cite{3c}.
{\it On the right}: preliminary upper limits (90\% CL) for the searches of neutrinos in coincidence with Gamma Ray Bursts in AMANDA \cite{grbama}, IceCube-22 \cite{grb22} and IceCube-40.}
\label{fig:pred}
\end{center}
\end{figure}

Fig. \ref{fig:point} shows, on the left, the sensitivity (90\% CL) of IceCube for the full-sky search of steady point sources of E$^{-2}$ muon neutrinos as a function of declination. The extension of the point source search to the southern hemisphere is made possible by rejecting background events by five orders of magnitude with a high energy event selection. This makes the southern hemisphere still dominated by high energy muon bundles and it yields a poorer neutrino detection sensitivity because of the high energy selection. Nevertheless this opens IceCube to a full-sky coverage and provides a coverage complement to the neutrino telescopes in the Mediterranean. On the right of Fig. \ref{fig:point} is the sky-map of statistical significance from the full sky search of IceCube-40. No significant localized excess is observed. The sensitivity is to be interpreted as the median upper limit we expect to observe from individual sources across the sky. If we test specific sources (see left panel of Fig. \ref{fig:pred}) we see that the full IceCube (about twice as sensitive as IceCube-40) will be able to discover neutrinos from individual point sources in about 3-5 years, depending on the location in the sky.

\vskip 0.2cm

The search for neutrinos from transient galactic and extra-galactic sources is also being pursued. In particular, the right panel in Fig. \ref{fig:pred} shows the upper limits (90\% CL) for the model-dependent search of prompt neutrinos from GRBs in the northern hemisphere with AMANDA \cite{grbama}, IceCube-22 \cite{grb22} and IceCube-40. For each detector configuration, the list of GRBs detected during the corresponding physics runs were collected and the predicted neutrino flux calculated based on the $\gamma$ ray spectrum \cite{guetta}. The corresponding average neutrino spectrum was used to search for neutrinos detected within the so-called $T_{90}$ time window (i.e. the time in which 5\% to 95\% of the fluence is recorded). The right panel of Fig. \ref{fig:pred} also shows the Waxman-Bahcall (WB) predicted average spectrum from GRBs \cite{wbgrb} and the average GRB spectrum corresponding to the 2009-2008 time period of IceCube-40 physics runs. The preliminary IceCube-40 upper limit is below the WB spectrum, which indicates that IceCube is becoming very sensitive and could potentially discover neutrinos in coincidence with GRBs within the next few years.

\subsection{Search for dark matter}
\label{ssec:dm}

Non-baryonic cold dark matter in the form of weakly  interacting massive particles (WIMPs) is one of the 
most promising solutions to the dark matter problem \cite{rubin}. The minimal supersymmetric extension of the Standard Model (MSSM) provides a natural WIMP candidate in the lightest neutralino $\tilde{\chi}^0_1$ \cite{drees}. This particle is weakly interacting only and, assuming R-parity conservation, is stable and can therefore survive today as a relic from the Big Bang. A wide range of neutralino masses, m$_{\tilde{\chi}^0_1}$, from 46 GeV \cite{amsler} to a few tens of TeV \cite{gilmore} is compatible with observations and accelerator-based measurements. Within these bounds it is possible to construct models where the neutralino provides the needed relic dark matter density.

\vskip 0.2cm
\begin{figure}[h]
\begin{center}
\psfig{file=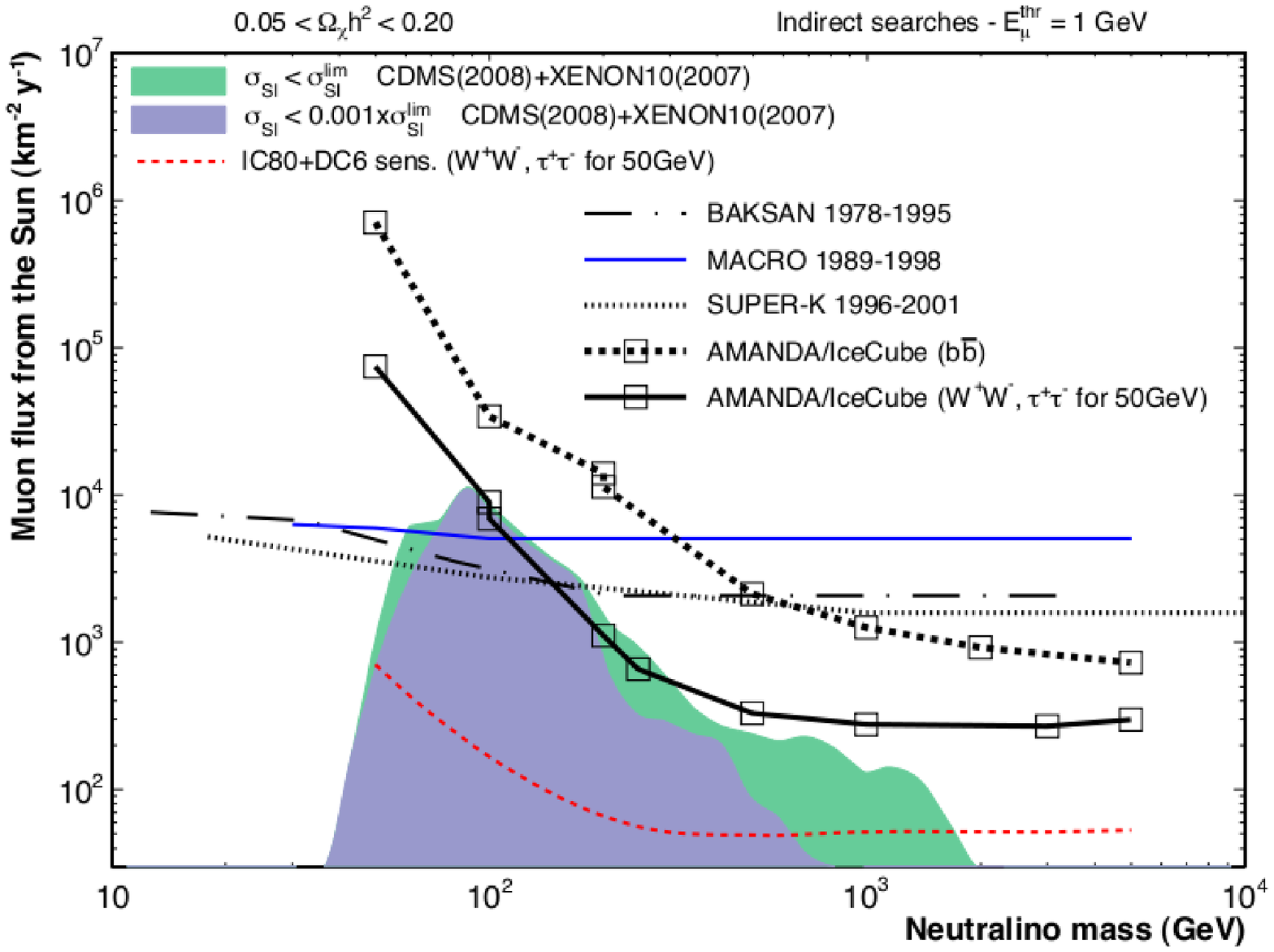,width=2.5in}
\psfig{file=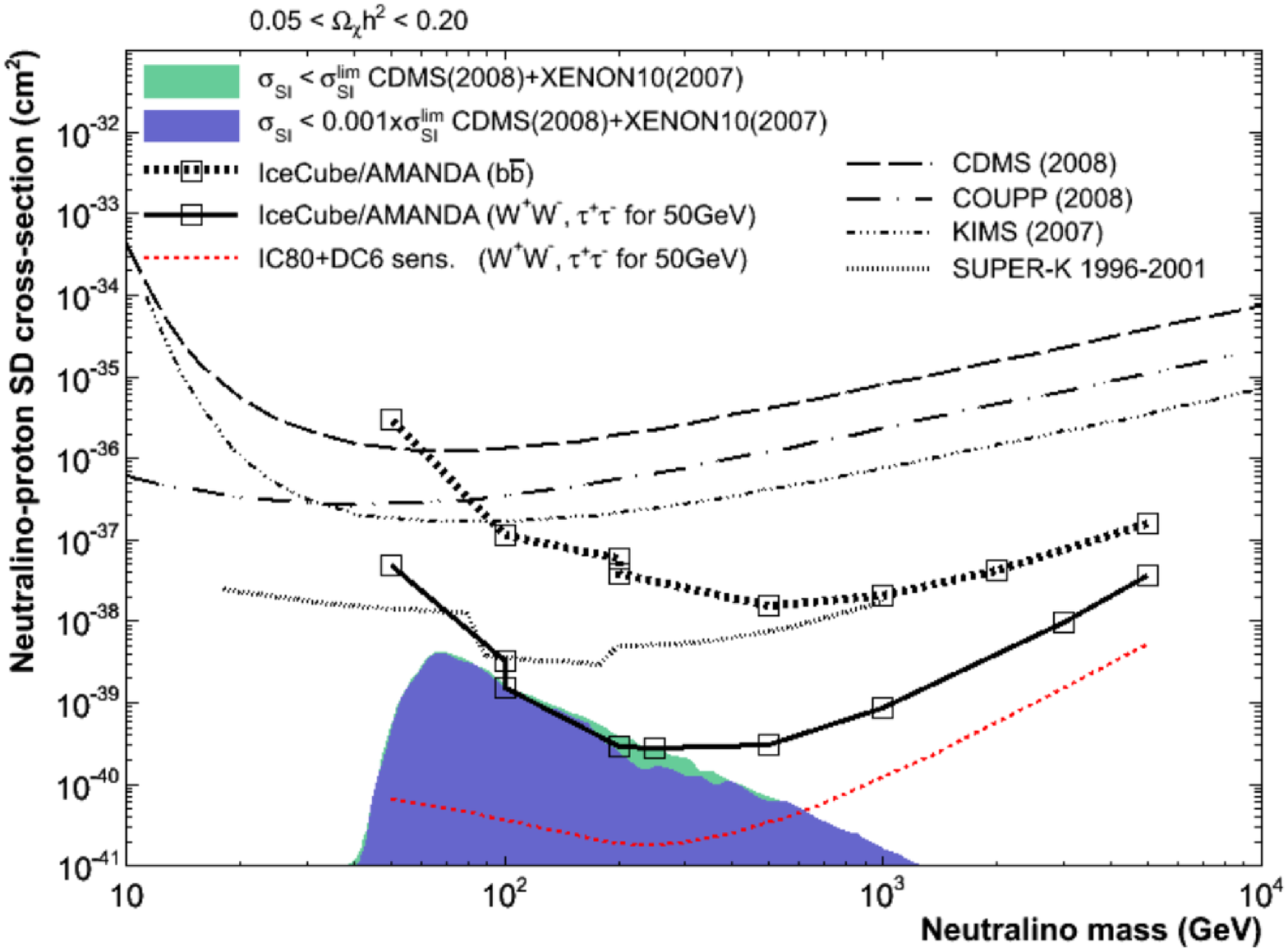,width=2.5in}
\caption{{\it On the left}: upper limits (90\% CL) on the muon flux from neutralino annihilations in the Sun for the soft ($b\bar{b}$) and hard (W$^+$W$^-$) annihilation channels, adjusted for systematic effects, as a function of neutralino mass. A muon energy threshold of 1GeV was used when calculating the flux. Also shown are the limits from MACRO \cite{ambrosio}, Super-K \cite{desai}, AMANDA \cite{ackermann}, and IceCube-22 \cite{ic22dm} merged to AMANDA at the low energy end. {\it On the right}: Upper limits (90\% CL) on the spin dependent neutralino-proton cross-section $\sigma^{SD}$ for the soft ($b\bar{b}$) and hard (W$^+$W$^-$) annihilation channels, adjusted for systematic effects, as a function of neutralino mass. Also shown are the limits from CDMS \cite{ahmed}, COUPP \cite{behnke}, KIMS \cite{lee} and Super-K \cite{desai}. The shaded area represents MSSM models not disfavored by direct searches \cite{ahmed, angle} based on $\sigma^{SI}$.}
\label{fig:dm}
\end{center}
\end{figure}

Relic neutralinos in the galactic halo may be gravitationally attracted by the Sun and accumulate in its center, where they can annihilate each other and produce standard model particles, such as neutrinos. This provides an indirect channel detection of this type of dark matter, provided the WIMP density and velocity distribution and the neutralino annihilation rate models are taken into account. The left panel of Fig. \ref{fig:dm} shows the upper limits (90\% CL) on the muon flux for IceCube-22 \cite{ic22dm} merged to the AMANDA upper limit at the low energy end \cite{ackermann}, along with other indirect observations. The limits on the annihilation rate can be converted into limits on the spin-dependent $\sigma^{SD}$ and spin-independent $\sigma^{SI}$ neutralino-proton cross-section (as shown on the right panel of Fig. \ref{fig:dm}). This conversion allows us a comparison with the direct search experiments. Since capture in the Sun is dominated by $\sigma^{SD}$ , indirect searches are expected to be competitive in setting limits on this quantity.  Assuming equilibrium between the capture and annihilation rates in the Sun, the annihilation rate is directly proportional to the cross-section. Fig. \ref{fig:dm} also shows the predicted sensitivity (90\% CL) of the combined IceCube-86 and the Deep Core dense instrumentation that allows us to significantly lower the energy threshold, and consequently increase sensitivity to low neutralino mass. In indirect searches WIMPs would accumulate in the Sun over a long period and therefore sample dark matter densities in a large volume of the galactic halo. This progressive gravitational accumulation is sensitive to low WIMP velocities, while direct detection recoil experiments are more sensitive to higher velocities, making indirect searches a good complement to the direct ones.

\begin{figure}[h]
\begin{center}
\psfig{file=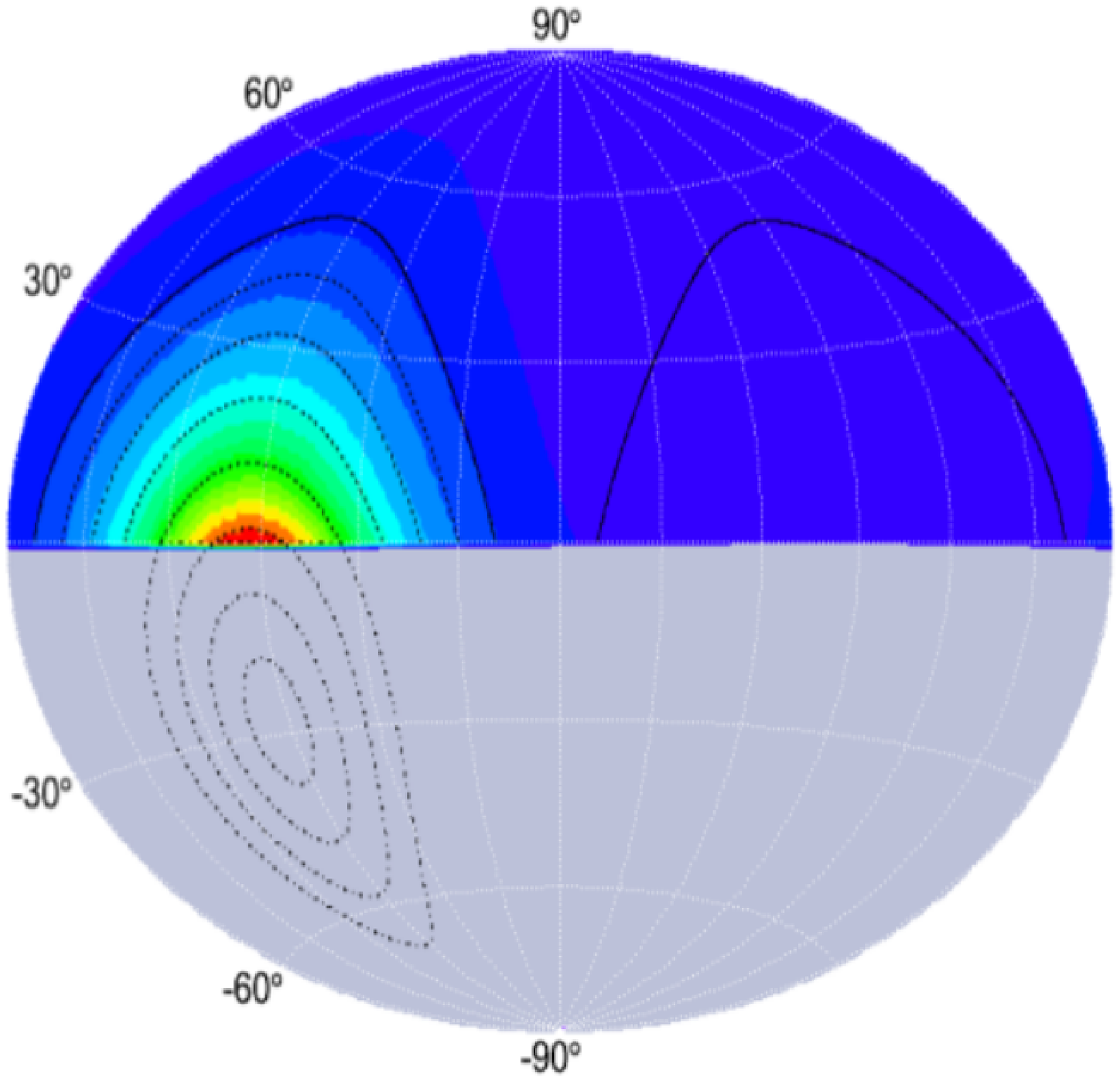,width=2.5in}
\psfig{file=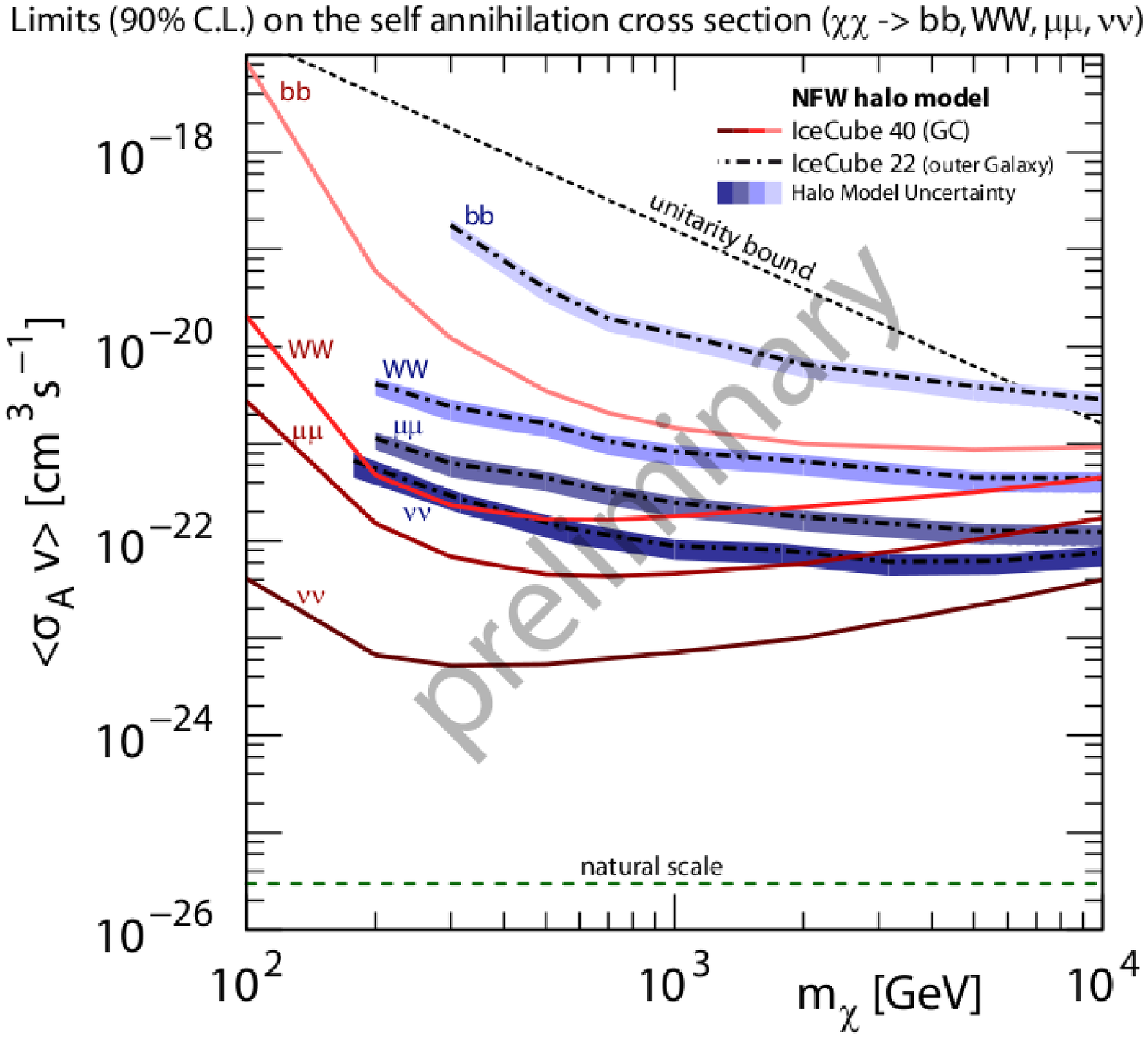,width=2.5in}
\caption{{\it On the left}:  relative expected neutrino flux from dark matter self-annihilations in the Milky Way 
halo on the northern celestial hemisphere (in equatorial coordinate). The largest flux is expected at right ascension $\alpha$ closest to the GC ($\Delta \alpha$ = 0). Dashed lines indicate circles around the Galactic center, while the solid lines show the definition of on and off-source region on the northern hemisphere. The on-source region is centered around $\Delta \alpha$ = 0, while the off-source region is rotated by 180$^{\circ}$ in $\alpha$. {\it On the right}: preliminary upper limits (90\% CL) on the self-annihilation cross-section $<\sigma_A v>$ as function of the WIMP mass m$_{\chi}$, for the search of an excess from the GC in the northern hemisphere with IceCube-22 \cite{carsten} and the search for down-ward neutrinos from the GC interacting inside the IceCube-40 array, for different annihilation channels. The bands on the IceCube-22 sensitivity curves account for different halo density profiles. The dashed line at the bottom labeled Ònatural scaleÓ is for dark matter candidates consistent with being a thermal relic. The black dashed line is the unitarity bound.}
\label{fig:halo}
\end{center}
\end{figure}

Neutrino telescopes can also test the dark matter self-annihilation cross section $<\sigma_A v>$ (averaged over the dark matter velocity distribution), making them complementary to $\gamma$ ray measurements. If the lepton excess observed by Fermi \cite{fermi2}, H.E.S.S. \cite{hess} and PAMELA \cite{pamela} is interpreted as a dark matter self-annihilation signal in the galactic dark matter halo \cite{meade}, the leptophilic dark matter in the TeV mass range provides the most compatible model. Since the dark matter halo column density is larger toward the direction of the Galactic Center (GC), neutrinos from WIMP self-annihilation in the halo are expected to have a large angular scale anisotropy with an excess in the direction of the GC (see left panel of Fig. \ref{fig:halo}). The dark matter density distribution in the Milky Way has different shapes depending on the model \cite{carsten}. The expected neutrino flux from dark matter self-annihilation is proportional to the square of the dark matter density integrated along the line of sight for a given angular distance from the Galactic center. The differential neutrino flux for a WIMP of mass m$_{\chi}$ depends on the halo density profile, the neutrino production multiplicity, and on self-annihilation cross section $<\sigma_A v>$ \cite{yuksel}. The search for an excess of neutrinos in the direction of the GC for different annihilation channels (assuming 100\% branching ratio for each of them), allows us to probe the allowed range of $<\sigma_A v>$ for the corresponding channels in a model-independent manner (see right panel of Fig. \ref{fig:halo}), and provide a direct comparison to similar results from $\gamma$ ray experiments.

\vskip 0.2cm

IceCube's reach can be significantly improved by directly looking at the GC (visible from the southern hemisphere). This search was performed with the IceCube 40-string dataset, by using down-ward neutrinos interacting inside the detector volume, and the preliminary upper limits (90\% CL) on the self-annihilation cross-section are shown on the right of Fig \ref{fig:halo}. While such an analysis is already able to put significantly better constraints, a large scale anisotropy would provide a more distinct discovery signal.

\subsection{Cosmic ray anisotropy}
\label{ssec:anyso}

Galactic cosmic rays are found to have an energy dependent large angular scale anisotropy in arrival direction distribution with amplitude of about $10^{-4}-10^{-3}$. The first comprehensive observation of such an anisotropy was provided by a network of muon telescopes sensitive to sub-TeV cosmic ray energies and located at different latitudes \cite{nagashima, hall}. More recently, an anisotropy was also observed in the multi-TeV range by the Tibet AS$\gamma$ array \cite{amenomori}, ARGO YBJ \cite{argo}, Super-K \cite{guillian}, and by MILAGRO \cite{abdo2}. And the first observation in the southern hemisphere is being reported by IceCube \cite{anisotropy} for a median cosmic ray energy of about 20 TeV.

\vskip 0.2cm
\begin{figure}[h]
\begin{center}
\psfig{file=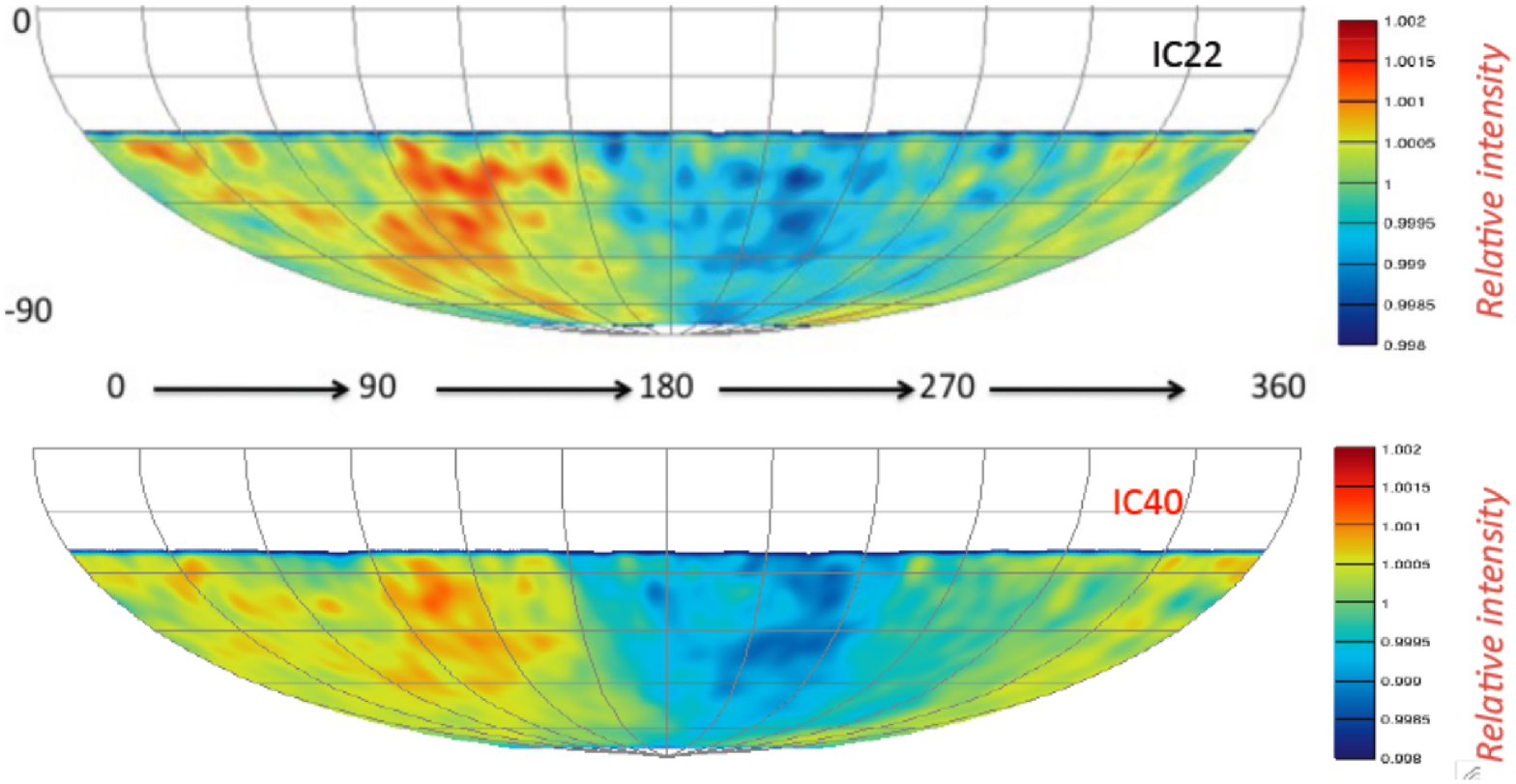,width=2.9in}
\psfig{file=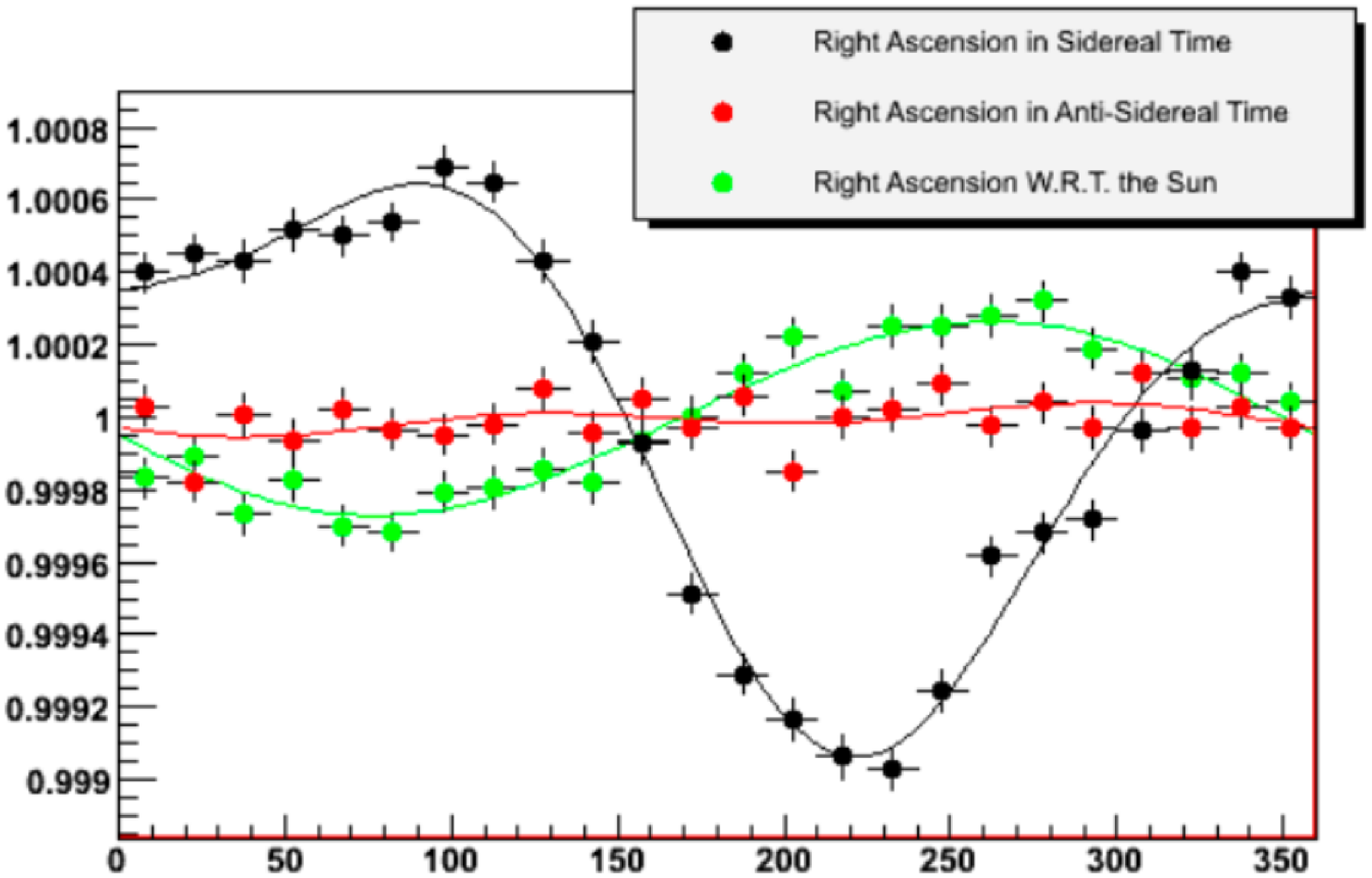,width=2.2in}
\caption{{\it On the left}: sky-map of the relative intensity in arrival direction of cosmic rays for IceCube-22 (on the top) \cite{anisotropy}, and preliminary sky-map of the relative intensity for IceCube-40 (on the bottom), in equatorial coordinates. For a better visual effect, a 3$^{\circ}$ smoothing has been applied to the maps. {\it On the right}: preliminary modulation of the relative intensity in arrival direction of cosmic rays for IceCube-40, projected in right ascension (black symbols), in the right ascension with respect to the Sun's location (green symbols), and in pseudo-right ascension, i.e. corresponding to anti-sidereal time (red symbols).}
\label{fig:anisotropy}
\end{center}
\end{figure}

The left panel of Fig. \ref{fig:anisotropy} shows the relative intensity in arrival direction of the cosmic rays, obtained by normalizing each $\sim$3$^{\circ}$ declination band independently. On the top is the relative intensity map obtained from the 4.6 billion events collected by IceCube-22 \cite{anisotropy} and on the bottom the preliminary map obtained from the 12 billion events collected by IceCube-40. The two maps show the same anisotropy features and they both appear to be a continuation of the observed modulation in the northern hemisphere. The right panel of Fig. \ref{fig:anisotropy} shows the preliminary modulation of the relative intensity in arrival direction of the cosmic rays projected into right ascension for IceCube-40 (black symbols). In order to verify whether the observed sidereal anisotropy (i.e. in equatorial coordinates), has, in one full year, some spurious modulation derived from the interference between possible yearly-modulated daily variations, the same analysis was performed using the anti-sidereal time frame (a non-physical time defined by switching the sign of the transformation from universal to sidereal time) \cite{farley}. The real feature in the sidereal time is expected to be scrambled in the anti-sidereal time. The anti-sidereal modulation (shown on the right of Fig. \ref{fig:anisotropy} in red symbols) appears to be relatively flat with an amplitude that is the same order of magnitude of statistical errors, suggesting that no significant spurious effect is present. If the relative intensity is measured, over one full year, as a function of the angular distance from the Sun in right ascension, we expect an excess in the direction of motion of the Earth around the Sun (at $\sim$ 270$^{\circ}$) and a minimum in the opposite direction. This is what is observed (see the green symbols on the right of Fig. \ref{fig:anisotropy}).

\vskip 0.2cm
\begin{figure}[h]
\begin{center}
\psfig{file=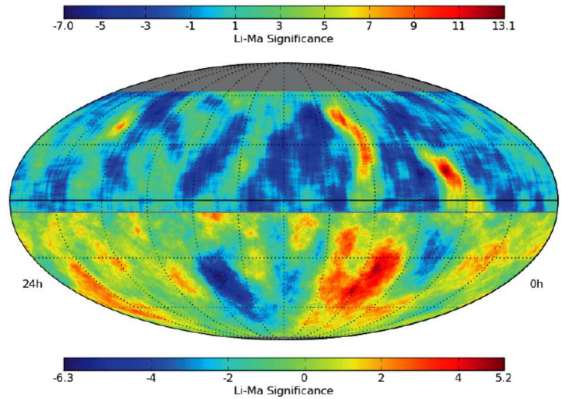,width=3in}
\caption{Preliminary statistical significance sky-map of the medium scale anisotropy in arrival direction of the cosmic rays for IceCube-40, combined with the MILAGRO northern hemisphere sky-map \cite{abdo3}. In this map only anisotropy scales that are smaller than $\sim$30$^{\circ}$ are visible. Note the different color scales for the statistical significance in the two hemispheres.}
\label{fig:residual}
\end{center}
\end{figure}

With the same techniques used in $\gamma$ ray detection, it is possible to estimate the event intensity sky-map (the background) with a pre-defined angular scale averaging, and determine the residual by subtracting the background from the actual map. With this method MILAGRO, in an attempt to estimate the background without $\gamma$-hadron separation, discovered two significant localized regions of cosmic rays \cite{abdo3}, also observed by ARGO YBJ \cite{vernetto}. The same medium-scale anisotropy measurement was performed with IceCube for the first time in the southern hemisphere and the combined MILAGRO-IceCube-40 significance sky-map is shown in Fig. \ref{fig:residual}, where only the anisotropy features with angular scale smaller than $\sim$30$^{\circ}$ are visible. The different event statistics between MILAGRO (with 220 billion events with median energy of $\sim$ 1 TeV) and IceCube-40 (with 12 billion events with median energy of $\sim$ 20 TeV) does not allow the comparison of the two hemispheres on a statistical base. Nevertheless, there seems to be some indication that the small scale features observed in the two hemispheres might be part of a larger scale structure.

The origin of the galactic cosmic ray anisotropy is still unknown. However there might be multiple superimposed causes, depending on the cosmic ray energy and the angular scale of the anisotropy. The large scale structure in the 10-100 TeV range might be a local fluctuation caused by a nearby supernova (within 1,000 pc) that exploded within the last 100,000 years or so \cite{erlykin}. On the other hand, the structure of the local Inter-Stellar Medium (ISM) magnetic field well within 1 pc might likely have an important role \cite{battaner}. The strongest and most localized of the MILAGRO excess regions has triggered astrophysical interpretations, invoking Geminga pulsar as a possible source \cite{astro1, astro2}. A strong anisotropy of the Magneto-Hydro Dynamic turbulence in the ISM, could cause a superposition of the large scale anisotropy (perhaps generated by a nearby SNR) with a beam of cosmic rays focused along the local magnetic field direction, depending on the turbulence scale \cite{astro3}. However, the localized nature of the hottest MILAGRO excess region suggests a local origin. Neutron monitor data also seem to indicate that the broad excess toward the direction of the heliotail\footnote{The part of the heliosphere opposite to the direction of the interstellar wind} (the so-called tail-in excess), which includes the MILAGRO localized excess regions, is likely generated within the heliotail itself \cite{neutron}. In particular the tail-in excess and its small angular scale structure, is suggestive of acceleration via magnetic reconnection in the solar magnetotail. Reconnection is generated by magnetic polarity reversals due to the 11-year solar cycles compressed by the solar wind in the magnetotail. The maximum energy of protons that can be accelerated through this process is estimated to be about 10 TeV. Up to this energy scale a localized excess might be observable in the direction of the acceleration sites \cite{helio}. This is the energy at which MILAGRO observes a cut-off for the localized regions.

\section{Acknowledgements}

{\small \small We acknowledge the support from the following agencies: U.S. National Science Foundation-Office of Polar Programs, U.S. National Science Foundation-Physics Division, University of Wisconsin Alumni Research Foundation, U.S. Department of Energy, and National Energy Research Scientific Computing Center, the Louisiana Optical Network Initiative (LONI) grid computing resources; Swedish Research Council, Swedish Polar Research Secretariat, Swedish National Infrastructure for Computing (SNIC), and Knut and Alice Wallenberg Foundation, Sweden; German Ministry for Education and Research (BMBF), Deutsche Forschungsgemeinschaft (DFG), Research Department of Plasmas with Complex Interactions (Bochum), Germany; Fund for Scientific Research (FNRS-FWO), FWO Odysseus programme, Flanders Institute to encourage scientific and technological research in industry (IWT), Belgian Federal Science Policy Office (Belspo); Marsden Fund, New Zealand; Japan Society for Promotion of Science (JSPS); the Swiss National Science Foundation (SNSF), Switzerland; A. Kappes and A. Gro\ss~acknowledges support by the EU Marie Curie OIF Program; J. P. Rodrigues acknowledges support by the Capes Foundation, Ministry of Education of Brazil.}

\end{document}